\documentclass[aps,prl,floatfix,epsfig,twocolumn,showpacs,preprintnumbers]{revtex4}
\usepackage{amssymb}
\usepackage{eurosym}
\usepackage{amsmath}
\usepackage{graphicx}
\usepackage{epstopdf}
\usepackage{dcolumn}
\usepackage{bm}
\usepackage{mathrsfs}
\usepackage[pdfstartview=FitH, CJKbookmarks=true, bookmarksnumbered=true, bookmarksopen=true, pdfborderstyle={/S/U}, pdfborder={0 0 1}, colorlinks=true, linkcolor=blue, anchorcolor=green, citecolor=blue]{hyperref}

\def \be{\begin{equation*}}
\def \ee{\end{equation*}}

\DeclareMathAlphabet {\mathbf}{OT1}{cmr}{bx}{n}

\begin{document}

\title{Calculated Exchange Interactions and Sensitivity of Ni Two--Hole Spin
State to Hund's Coupling in Doped NdNiO$_{2}$ }
\author{Xiangang Wan$^{\ast }$, Vsevolod Ivanov$^{\dag }$, Giacomo Resta$%
^{\dag }$, Ivan Leonov$^{\dag \dag }$, Sergey Y. Savrasov$^{\dag }$}
\affiliation{$^{\ast }$National Laboratory of Solid State Microstructures, School of
Physics and Collaborative Innovation Center of Advanced Microstructures,
Nanjing University, 210093 Nanjing }
\affiliation{$^{\dag }$Department of Physics and Astronomy, University of California,
Davis, CA 95616}
\affiliation{$^{\dag \dag }$M.N. Miheev Institute of Metal Physics, Russian Academy of
Sciences and Ural Federal University, 620002 Yekaterinburg}

\begin{abstract}
Using density functional based LDA+U\ method and linear--response theory, we
study the magnetic exchange interactions of the superconductor Nd$_{1-x}$Sr$%
_{x}$NiO$_{2}$. Our calculated nearest-neighbor exchange constant $J_{1}=82$
meV is large, weakly affected by doping and is only slightly smaller than
that found in the sister compound CaCuO$_{2}.$ We however find that the hole
doping significantly enhances the inter--layer exchange coupling as it
affects the magnetic moment of the Ni--3\emph{d}$_{3z^{2}-r^{2}}$ orbital.
This can be understood in terms of small hybridization of Ni-3\emph{d}$%
_{3z^{2}-r^{2}}$ within the NiO$_{2}$ plane which results in a flat band
near the Fermi level, and its large overlap along z direction. We also
demonstrate that the Nd-5\emph{d} states appearing at the Fermi level, do
not affect the magnetic exchange interactions, and thus may not participate
in the superconductivity of this compound. Whereas many previous works
emphasize the importance of the Ni-3\emph{d}$_{x^{2}-y^{2}}$ and Nd-5\emph{d}
orbitals, we instead analyze the solution of Ni-3\emph{d}$_{x^{2}-y^{2}}$%
/Ni-3\emph{d}$_{3z^{2}-r^{2}}$ minimal model using Dynamical Mean Field
Theory. It reveals an underlying Mott insulating state which, depending on
precise values of the intra--atomic Hund's coupling less or larger than 0.83
eV, selects upon doping either S=0 or S=1 two--hole states at low energies
leading to very different quasiparticle band structures. We propose that
trends upon doping in spin excitational spectrum and quasiparticle density
of state can be a way to probe Ni 3d$^{8}$ configuration.
\end{abstract}

\date{\today }
\maketitle

\section{I. \textbf{INTRODUCTION}}

Since the discovery of high--temperature superconductors (HTSCs)\cite{HTC},
tremendous theoretical and experimental efforts have been devoted to
understanding the novel physics of this family of compounds\cite%
{HTC-1,HTC-2,HTC-3}. All HTSCs are comprised of quasi--two--dimensional CuO$%
_{2}$ planes separated by charge reservoir spacer layers, and their parent
compounds have antiferromagnetic (AFM) order with very strong in--plane
magnetic exchange interactions, belonging to the class of charge--transfer
insulators\cite{ZSA}. Upon doping, holes occupy the O--2\emph{p} orbital,
and due to the strong hybridization between Cu--3\emph{d}$_{x^{2}-y^{2}}$
and O--2\emph{p} orbitals, a Zhang--Rice singlet is formed\cite{Zhang-Rice}.
It has been widely accepted that the HTSCs can be described by an effective
single band t--J model, with different parameters explaining the variation
in $T_{c}$ in different materials \cite{t-J-1,t-J-2}.

Inspired by HTSCs, the search for possible novel superconducting behavior in
nickelates has been attracting significant attention, as their structure and
electronic configuration is similar to that of the cuprates \cite%
{LaNiO2-exp-JPCS-1996,LaNiO2-exp-JAC-1998,LaNiO2-exp-JACS-1999,LaNiO2-exp-Phys.C.-2013}%
. Unfortunately, the monovalent Ni ion is strongly unstable and scarcely
formed in mineral compounds, making, for example, LaNiO$_{2}$ difficult but
possible to synthesize\cite{LaNiO2-exp-JAC-1998}. First--principles Local
Density Approximation (LDA) based calculations revealed an important
difference between LaNiO$_{2}$ and its sister infinite--layer HTSC compound
CaCuO$_{2}$ : the Fermi surface of CaCuO$_{2}$ consists of only one
two--dimensional band, while LaNiO$_{2}$ seems quite three--dimensional,
with La--derived 5\emph{d} states and Ni--3\emph{d} states crossing the
Fermi level\cite{Pickett-2004}. Numerical calculations predicted AFM
magnetic order\cite{Pickett-2004,LaNiO2-LDA+U-1999}, but magnetization and
neutron powder diffraction observe no long--range order in LaNiO$_{2}$\cite%
{LaNiO2-exp-JACS-1999}. At high temperatures (150K $<$ T $<$ 300K), the
susceptibility of LaNiO$_{2}$ can be fitted by the sum of a temperature
independent term, and a Curie--Weiss $S=\frac{1}{2}$ paramagnetic term with
a large Weiss constant ($\theta =-257$K), indicating a significant
correlation between Ni spins\cite{LaNiO2-exp-JACS-1999}. LaNiO$_{2}$ shows
metallic behavior, but resistivity increases at lower temperatures and no
superconducting state has been observed \cite%
{LaNiO2-exp-JPCS-1996,LaNiO2-exp-JAC-1998,LaNiO2-exp-JACS-1999}.

Recently, Nd$_{0.8}$Sr$_{0.2}$NiO$_{2}$ thin films were synthesized on a
SrTiO$_{3}$ substrate using soft--chemistry topotactic reduction, and
superconductivity with considerably high T$_{c}$ (up to 15K) was observed%
\cite{Nature-2019}. The superconducting phase displays a doping--dependent
dome for Nd$_{1-x}$Sr$_{x}$NiO$_{2}$ (0.125 $<$ x $<$ 0.25), which is
remarkably similar to that of the cuprates \cite%
{Superconducting-Dome-1,Superconducting-Dome-2}. Very recently,
superconductivity has also been observed in doped PrNiO$_{2}$\cite{PrNiO2}.

These breakthroughs have stimulated\ large--scale theoretical efforts to
understand the nature of the superconductivity in rare--earth nickelates.
LDA band structures \cite{Norman} predict that both Nd--5\emph{d} and Ni--3%
\emph{d}$_{x^{2}-y^{2}}$ orbitals contribute significantly to the Fermi
surface of parent compound NdNiO$_{2}$. Most calculations treat the three 4%
\emph{f} electrons in Nd$^{3+}$ as core electrons, although the role of Nd--4%
\emph{f} has been emphasized recently\cite{Pickett-2}. Many--body
perturbative GW calculations result in almost no modification to the
Fermi--surface topology and its orbital composition\cite{GW}. Focusing on
the Fermi surface, different minimal models have been proposed to describe
the low energy physics of this material using a Wannier function approach,
including: a three--band model with Ni-3\emph{d}$_{x^{2}-y^{2}}$, Nd--5\emph{%
d}$_{3z^{2}-r^{2}}$ and an interstitial \emph{s} orbital\cite{Arita-2019}; a
three--band model with Ni--3\emph{d}$_{x^{2}-y^{2}}$, Nd--5\emph{d}$%
_{3z^{2}-r^{2}}$ and Nd--5\emph{d}$_{xy}$\cite{J.L.Yang,H.M.Weng}; a
two--band model with Ni--3\emph{d}$_{x^{2}-y^{2}}$ and Nd--5\emph{d}$%
_{3z^{2}-r^{2}}$\cite{Nature-Mater-2020,Wannier}; and a four--band model\cite%
{H.H.Chen}. The effect of topotactic hydrogen has been discussed as well\cite%
{Held}.

Several works addressed strong correlation effects among Ni 3d electrons 
\cite{Sawatzky}--\cite{Spin excitations} Due to a large energy difference
between O--2\emph{p} and Ni--3\emph{d} levels, the undoped NdNiO$_{2}$ has
been suggested to be a Mott insulator, and a coexistence/competition between
low energy S=0 and S=1 states has been proposed for the hole doped case \cite%
{Sawatzky} where some Ni ions would acquire a formal 3$d^{8}$ configuration.
The origin of these two--hole states has been discussed in a recent
literature\cite%
{Lechermann,C.J.Wu,t-J-Ashvin,Millis,Multiorbital,Multiorbital2}. As it is
commonly accepted that the undoped Ni 3$d^{9}$ configuration corresponds to
the hole of $x^{2}-y^{2}$ symmetry, the two--hole states produced by doping
can either end up as intraorbital singlets or interorbital triplets. It has
been first pointed out\cite{Sawatzky} that the S=1 state maybe incompatible
with robust superconductivity, and indeed exact diagonalization study of Ni
impurity embedded into the oxygen environment \cite{Sawatzky} as well as a
number of many--body calculations using a combination of LDA with Dynamical
Mean Field Theory (DMFT)\cite{Lechermann,Millis} pointed to the formation of
the intraorbital singlets.

The first--principles physics of competing Ni-3\emph{d}$_{x^{2}-y^{2}}$ vs.
Ni--3\emph{d}$_{3z^{2}-r^{2}}$, and the connected issue of having the Ni--3%
\emph{d}$_{3z^{2}-r^{2}}$ states at the Fermi level with hole doping has
been first put forward in Ref. \cite{Multiorbital}. In addition, the role of
Ni-3\emph{d}$_{x^{2}-y^{2}}$ and Ni-3\emph{d}$_{3z^{2}-r^{2}}$ orbitals has
been emphasized in Ref. \cite{Lechermann}. A recent GW+DMFT work \cite%
{GW+DMFT}, highlighted Ni-3\emph{d}$_{3z^{2}-r^{2}}$ flat--band physics as
well as Ref. \cite{Pickett-3}. Furthermore, a variant of the t--J model with
S=1 has been proposed and shown to exhibit d--wave superconductivity\cite%
{t-J-Ashvin}. Symmetries of the pairing states based on a two--orbital Ni-3%
\emph{d}$_{x^{2}-y^{2}}$/Ni-3\emph{d}$_{xy}$ model Hamiltonian with
competing S=0 and S=1 two--hole states have been discussed\cite{C.J.Wu}.
DMFT calculations for the two--orbital Ni--3\emph{d}$_{x^{2}-y^{2}}$/Ni--3%
\emph{d}$_{3z^{2}-r^{2}}$ system argued that a multiorbital description of
nickelate superconductors is necessary \cite{Multiorbital2}. Excitations and
superconducting instabilities have also been explored by a random phase
approximation \cite{Wannier+RPA}and by a variant of the t-J model\cite%
{effective model}. Local spin, charge and orbital susceptibilities have been
calculated using a combination of DMFT\ with a local quasiparticle
self--consistent GW method and emphasized the Hund's physics of Ni--$e_{g}$
electrons\cite{kotliar}.

No sign of magnetic order has been observed in the original report on
superconductivity in NdNiO$_{2}$\cite{Nature-2019}, which may be attributed
to defects, such as unwanted hydrides or hydroxides that might form as
by--products of the creation of the rare Ni$^{+}$ oxidation state during the
synthesis of this compound. Another consideration is that LaNiO$_{3}$ is
close to an antiferromagnetic quantum critical point (QCP)\cite%
{NC-LaNiO3-2020}, therefore it is reasonable to expect that with lower
dimensionality, NdNiO$_{2}$ would pass the QCP and display magnetism. Very
recently, strong spin fluctuations and considerable AFM exchange
interactions have been observed in NdNiO$_{2}$\cite{Raman J.W.Mei} as well
as nuclear magnetic resonance (NMR) data \cite{NMR} provided an addtional
evidence for quasi--static AFM order below 40 K and dominant spin
fluctuations at higher temperatures in Nd$_{0.85}$Sr$_{0.15}$NiO$_{2}$ bulk
materials. The exchange interactions have also been discussed in several
works \cite{J.L.Yang,J.W.Mei,M.J.Han,Arita-J}. The calculated
electron--phonon interaction ($\lambda \leq $ 0.32) is too small to explain
the 15K T$_{c}$ in this material\cite{Arita-2019}, meaning the spin
excitations, which are thought to be responsible for the superconductivity
in HTSCs\cite{HTC-1,HTC-2,HTC-3}, are worth careful investigation.

In this work, based on a density functional LDA+U\ method and
linear--response theory\cite{J-method}, we perform detailed studies of
exchange interactions for both parent and doped NdNiO$_{2}$. The method does
not rely on a total energy analysis, and instead directly computes the
exchange constant for a given wave vector $\mathbf{q}$ based on the result
of the magnetic force theorem \cite{Magnetic-force-theroem}. Our results
show that although the Fermi surface of undoped NdNiO$_{2}$ is quite
three--dimensional, its magnetic exchange interaction $J$ has a clear
two--dimensional feature with large in--plane $J_{1}=82$ meV and much
smaller out--of--plane $J_{z1}$. However, the Ni--3\emph{d}$_{3z^{2}-r^{2}}$
band close to the Fermi level is quite flat, therefore within the LDA+U\
method for a reasonable range of the values of Hubbard $U$ above 4 eV, holes
introduced by doping preferentially occupy the Ni--3\emph{d}$_{3z^{2}-r^{2}}$
orbitals while Ni--$t_{2g}$ states remain remarkably inert. The in--plane $%
J_{1}$ remains largely unaffected by doping, but the magnetic moment of the
Ni--3\emph{d}$_{3z^{2}-r^{2}}$ orbital and the out--of--plane $J_{z1}$ both
grow significantly in accord with recent findings\cite{Pickett-3} Our
calculation using a constrained--orbital--hybridization method \cite%
{shift-orbital} unambiguously demonstrates that while Nd--5\emph{d} makes an
important contribution to the Fermi surface, it has almost no effect on the
magnetic exchange interaction. It is expectable result, since it is known
that Nd-5d orbitals have negligible hybridization with Ni orbitals \cite%
{Arita-2019,ZC Zhong}. This means the magnetic excitations in hole--doped
NdNiO$_{2}$ can be described by an effective model including Ni--3\emph{d}$%
_{x^{2}-y^{2}}$/Ni--3\emph{d}$_{3z^{2}-r^{2}}$ orbitals whose role has been
emphasized in many recent works \cite{Norman}--\cite{Spin excitations}.

To gain additional insight, we discuss the solutions of such two--band model
on the basis of Dynamical Mean Field Theory using the parameters deduced
from our band structure calculations. In contrast to the static mean field
description, such as LDA+U, where holes occupying Ni--3\emph{d}$%
_{3z^{2}-r^{2}}$ states promote interorbital triplets, whether S=0 or S=1
state emerges from our DMFT simulation depends on a precise value of the
intraatomic Hund's coupling $J_{H}$ in the vicinity of its commonly accepted
range of values 0.5--1 eV. This leads to very different quasiparticle band
structures. We thus propose that trends upon doping in magnetic exchange
interactions and quasiparticle density of states can be a way to probe Ni 3d$%
^{8}$ configuration.

Our paper is organized as follows. In Section II we describe our LDA+U and
constrained--orbital--hybridization calculations for the exchange
interactions. In Section III, we discuss a minimal two--band model that
emerges from our study and its solution based on Dynamical Mean Field
Theory. Section IV is the conclusion.

\begin{figure}[tbp]
\includegraphics[width=0.49\textwidth]{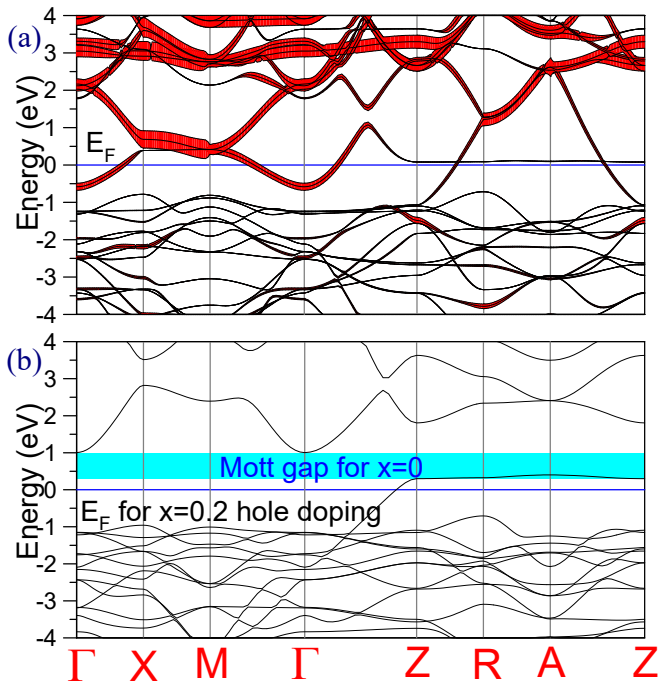}
\caption{Band structure of ($\protect\pi $,$\protect\pi $,0) AFM ordered
NdNiO$_{2}$ from LDA+$U$ calculations with $U=6.0$ eV. (a) undoped NdNiO$%
_{2} $ with Nd--5$d$ oribtal character shown in red, (b) constrained
orbital-hybridization calculation for NdNiO$_{2}$ with the Nd--5d band
shifted up by $2$ Ry. The position of the Fermi level corresponds to 0.2
hole doping.}
\label{FigBands}
\end{figure}

\section{II. \textbf{CALCULATIONS\ OF\ EXCHANGE\ INTERACTIONS}}

We perform our density functional based electronic structure calculations
within the full potential linear--muffin--tin--orbital (LMTO) method\cite%
{FP-LMTO}. To take into account the effect of on--site electron--electron
interactions between Ni--3\emph{d} orbitals we add a correction due to
Hubbard $U$ using the so--called LDA+$U$ approach\cite{LDA+U}. Although, the
experimental situation on magnetism in nickelates is still unclear, hinted
by the cuprate physics, an assumption of the AFM ordered state in the parent
compound should be a good starting point for a theoretical modeling.
Ultimately, if AFM spin fluctuations in the doped state are responsible for
superconductivity, the exchange interactions in the ordered state set the
scale for those fluctuations, which justifies this assumption and provides
the basis for our static linear response calculation of J's. An alternative
measure of those spin fluctuations would be a full calculation of wavevector
and frequency dependent spin susceptibility directly in paramagnetic state.
Although possible, in principle, it is a lot more involved and goes beyond
the scope of this work.

We vary the parameter $U$ for Ni--3\emph{d} between 4.0 and 8.0 eV, and find
that the essential properties and our conclusions do not depend on the value
of $U$ in this range\cite{SM}. Below we report our results for exchange
constants with $U=6$ eV and Hund's $J_{H}=0.95$ eV. Experimental lattice
parameters have been used\cite{Nature-2019}.

The magnetic exchange interactions $J(\mathbf{q})$ were evaluated assuming a
rigid rotation of atomic spins, using a previously developed
linear--response approach \cite{J-method}. This technique has been applied
successfully to evaluate exchange interactions for a series of materials,
including transition--metal oxides\cite{J-method}, HTSCs\cite{HTC-J},
Fe--based superconductors\cite{Fe-based}; europium monochalcogenides\cite%
{EuO-J}, orbital--ordered noncollinear spinel MnV$_{2}$O$_{4}$\cite{MnV2O4},
and Dirac magnon material Cu$_{3}$TeO$_{6}$\cite{Cu3TeO6}. We also use a
constrained--orbital--hybridization method to provide theoretical insights
into the various contributions\cite{shift-orbital} to the exchange
interactions in hole doped NdNiO$_{2}$. To avoid the effect of the very
narrow Nd--4\emph{f} bands, we shift the three occupied Nd--4\emph{f}
orbitals downward while shifting the rest of the Nd--4\emph{f} band upward
by using a constrained--orbital approach \cite{shift-orbital}. Since the
obtained results do not depend on the magnitude of the shifts, we display
the results with the Nd--4\emph{f} bands shifted by $\pm 2.0$ Ry.

\begin{table}[tbp]
\caption{Calculated exchange interactions, $J_{1}$, $J_{2}$ (in-plane
nearest, next-nearest) and $J_{z1}$, $J_{z2}$ (out-of-plane nearest,
next-nearest) in meV for various hole dopings $x$. Positive/negative sign
denotes AFM/FM interaction. We also list the calculated total magnetic
moment at the Ni site, $M_{tot}$, and magnetic moment at the Ni-3\emph{d}$%
_{3z^{2}-r^{2}}$ orbital, $M_{3z^{2}-r^{2}}$, (in $\protect\mu _{B}$). }
\label{doping}%
\begin{tabular}{c|cccccc}
\hline\hline
$x$ & $J_{1}$ & $J_{2}$ & $J_{z1}$ & $J_{z2}$ & $M_{tot}$ & $%
M_{3z^{2}-r^{2}} $ \\ \hline
0.00 & 82.24 & -4.84 & -3.40 & -23.00 & 0.97 & 0.17 \\ 
0.05 & 71.84 & -5.08 & -21.40 & -22.40 & 1.03 & 0.23 \\ 
0.10 & 65.88 & -5.68 & -39.36 & -18.48 & 1.07 & 0.27 \\ 
0.15 & 64.60 & -4.68 & -59.04 & -11.08 & 1.08 & 0.30 \\ 
0.20 & 58.20 & -4.16 & -80.88 & -5.56 & 1.15 & 0.35 \\ 
0.25 & 57.36 & -2.84 & -97.36 & -4.84 & 1.16 & 0.38 \\ 
0.30 & 50.76 & -2.16 & -105.52 & -2.88 & 1.23 & 0.44 \\ \hline\hline
\end{tabular}%
\end{table}

Similarly to previously reported band structure calculations for LaNiO$_{2}$%
\cite{Pickett-2004,Arita-2019,H.M.Weng,H.H.Chen}, there are two bands
crossing the Fermi level in the LDA band structure of NdNiO$_{2}$, with one
band primarily derived from the Ni--3\emph{d}$_{x^{2}-y^{2}}$ orbital and
the other consisting of predominantly Nd--5\emph{d} character. Just as with
LaNiO$_{2}$\cite{Pickett-2004,Arita-2019}, there is a gap between Ni--3\emph{%
d} and O--2\emph{p} bands (around -3.5 eV). Moreover, Ni--O bond length in
NdNiO$_{2}$ (1.96 $\mathring{A}$) is slightly larger than the Cu--O bond
length in CaCuO$_{2}$ (1.92 $\mathring{A}$). Thus the bandwidth of the Ni--3%
\emph{d}$_{x^{2}-y^{2}}$ band correspondingly smaller than that of the Cu--3%
\emph{d}$_{x^{2}-y^{2}}$ band. While in both NdNiO$_{2}$ and CaCuO$_{2}$,
the 3\emph{d}$_{3z^{2}-r^{2}}$ orbitals have very small dispersions along
the $ZRAZ$ line, the dispersion of the Ni--3\emph{d}$_{3z^{2}-r^{2}}$ state
along $\Gamma Z$ is considerably larger than that of Cu--3\emph{d}$%
_{3z^{2}-r^{2}}$. Moreover, compared to Cu--3\emph{d}$_{3z^{2}-r^{2}}$, the
Ni--3\emph{d}$_{3z^{2}-r^{2}}$ band lies closer to the Fermi level. These
two features are expected to significantly affect the magnetic behavior in
the hole doped NdNiO$_{2}$.

\begin{table}[tbp]
\caption{Calculated exchange interactions (in meV) for x=0.2 hole doped NdNiO%
$_{2}$, with Nd-5\emph{d} shifted upward by various energies (in Ry).}
\label{shift-5d}%
\begin{tabular}{c|cc}
\hline\hline
Shift (Ry): & $J_{1}$ & $J_{z1}$ \\ \hline
0.05 & 64.28 & -84.04 \\ 
0.10 & 65.48 & -83.76 \\ 
0.50 & 69.68 & -75.52 \\ 
2.00 & 71.92 & -78.96 \\ \hline\hline
\end{tabular}%
\end{table}

We now perform the LDA+$U$ calculation to examine magnetic exchange
interactions in undoped NdNiO$_{2}$. Our results show that the exchange
coupling is large for the nearest--neighbor $J_{1}$ within the NiO$_{2}$
plane. The sign of this term is AFM, and thus the NiO$_{2}$ layer shows a ($%
\pi $,$\pi $) spin ordering. There is some debate about the magnitude of the
exchange interaction, with estimates ranging from much less than that of
cuprates \cite{Sawatzky,J.L.Yang,J.W.Mei} to comparable to the value of
exchange interaction in CaCuO$_{2}$\cite{Wannier,M.J.Han,Arita-J}. Our
calculated value of $J_{1}\ $is 82.24 meV as referenced to the form of the
Heisenberg Hamiltonian%
\begin{equation}
H=\frac{1}{2}\sum_{ij}J_{ij}\mathbf{S}_{i}\mathbf{S}_{j}  \label{HH}
\end{equation}%
with $S$=1/2. The estimate of $J_{1}=$25 meV from the Raman scattering
experiment of the two--magnon peak \cite{J.W.Mei,Raman J.W.Mei} is
significantly smaller.\ Recent resonance X--ray scattering experiments
performed for trilayer nickelate La$_{4}$Ni$_{3}$O$_{8}$ report this value
to be 69 meV\cite{La4Ni3O8}. The in--plane $J_{1}$ that we compute is only
about 25\% less than that found in CaCuO$_{2}$\cite{HTC-J}. We attribute it
to a smaller Ni--3\emph{d} and O--2\emph{p} hybridization and larger energy
splitting between Ni--3\emph{d} and O--2\emph{p} as has previously been
pointed out\cite{Pickett-2004}. Consistent with the result of ($\pi $,$\pi $%
,0) spin ordering being slightly more energetically favorable than ($\pi $,$%
\pi $,$\pi $), our calculation produces a small out--of--plane FM exchange
interaction, with nearest neighbor $J_{1z}=-3.4$ meV and second nearest
neighbor $J_{2z}=-23$ meV, respectively. Our calculations reveal that the
magnetic moment at the Ni site (0.97 $\mu _{B}$), residing mostly in the 3d$%
_{x^{2}-y^{2}}$ orbital, is much larger than that at Cu sites in HSTCs.

There exists a fairly flat band right at the Fermi level along the $ZRAZ$
line, in the band structure of the magnetic ground state configuration of
NdNiO$_{2},$ as shown in Fig.\ref{FigBands}(a). This flat band has
predominantly Ni--3\emph{d}$_{3z^{2}-r^{2}}$ character, and plays an
important role when hole doping is considered. The very small in--plane
dispersion of the Ni--3\emph{d}$_{3z^{2}-r^{2}}$ band can be understood as a
consequence of the symmetry of the Ni--3\emph{d}$_{3z^{2}-r^{2}}$ orbital,
which can only weakly hybridize with the neighboring O--2\emph{p}.

To examine the doping dependence we perform a series of hole--doped
calculations, varying the number of holes per unit cell from 0.05 to 0.30 by
using the virtual crystal approximation. These calculations show that the
hole doping within this range does not significantly change the shape of the
band structure apart from shifting the Fermi level downward. Regardless of
the hole--doping concentration, the Ni--$t_{2g}$ band is almost fully
occupied and does not contribute to the magnetic moment. The magnetic moment
of the Ni--3\emph{d}$_{x^{2}-y^{2}}$ orbital is also unaffected by the hole
doping. Instead, the holes preferentially occupy the flat Ni--3\emph{d}$%
_{3z^{2}-r^{2}}$ band, and, as a result, the magnetic moment of this orbital
increases with doping as shown in Table \ref{doping}. Noting the
considerable Ni--3\emph{d}$_{3z^{2}-r^{2}}$ band dispersion along $\Gamma Z$%
, and the formation of magnetic moments in this orbital, one can expect the
emergence of out--of--plane magnetic exchange interactions. This result has
been confirmed by our linear response calculation. As shown in Table \ref%
{doping}, hole doping significantly enhances the out--of--plane $J_{z1}$,
while the in--plane $J_{1}$ remains mostly unaffected.

The 5\emph{d} orbital is spatially very wide, and can have a crucial effect
on the magnetic exchange interaction through 4\emph{f}--5\emph{d}
hybridization, even though it is empty and located above the Fermi level\cite%
{EuO-J}. In NdNiO$_{2}$, the Nd--5\emph{d} band appears at the Fermi level,
making it important to understand the role of the Nd--5\emph{d} orbital in
magnetic exchange interactions. We address this issue by using a
constrained--orbital--hybridization approach\cite{shift-orbital}. We perform
the calculations with the Nd--5\emph{d} band shifted upward by various
values. Fig. \ref{FigBands}(b) shows the band structure for the case where
the Nd--5\emph{d} band is shifted upward by 2 Ry. As one can see, the AFM
insulating state emerges from this calculation for the undoped case, while
hole doping vacates the Ni--3\emph{d}$_{3z^{2}-r^{2}}$ band within $%
k_{z}=\pi /c$ plane.

Our calculation shows that both in--plane $J_{1}$ and out--of--plane $J_{z1}$
exchange interactions are not sensitive to the position of the Nd--5\emph{d}
band as shown in Table \ref{shift-5d}, clearly indicating that the effect of
this orbital on the magnetic exchange interactions is negligible. A similar
calculation was performed for LaNiO$_{2}$ to further confirm these findings%
\cite{SM}. While the obtained values of the exchange interactions are
slightly different, the key features discussed above are the same.

\begin{figure}[tbp]
\includegraphics[height=0.233\textwidth,width=0.48\textwidth]{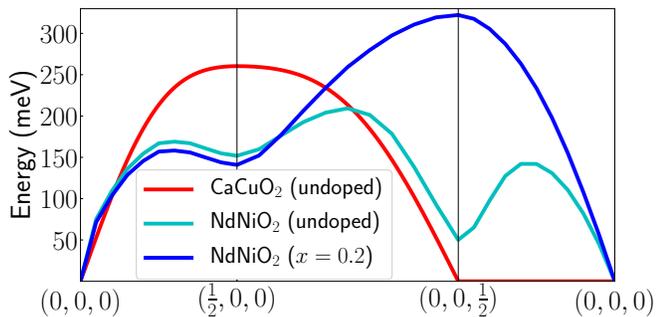}
\caption{Calculated spin--wave dispersions for the undoped NdNiO$_{2}$
(cyan), and 0.2 hole-doped NdNiO$_{2}$ (blue), with $U$=6eV. For comparison,
we also plot the results of CaCuO$_{2}$ (red)\protect\cite{HTC-J}.}
\label{FigSW}
\end{figure}

We illustrate the effect of increasing out--of--plane exchange interactions
in doped NdNiO$_{2}$, using an antiferromagnetic Heisenberg model, Eq. (\ref%
{HH}). Its linear spin--wave dispersion is given by%
\begin{equation*}
\omega (\mathbf{q})=S\sqrt{[J_{11}(\mathbf{q}%
)-J_{11}(0)+J_{12}(0)]^{2}-[J_{12}(\mathbf{q})]^{2}}
\end{equation*}%
where $J_{11}(\mathbf{q})$/$J_{12}(\mathbf{q})$ are the exchange
interactions within the same/different sublattices. (A quantum correction
factor $Z_{c}\approx 1.18$ which is sometimes used \cite{HTC-J} in front of
this formula is omitted here) We plot these dispersions in Fig. \ref{FigSW}
for both undoped and 0.2 hole--doped NdNiO$_{2}$ in Fig. \ref{FigSW}, along
with those of CaCuO$_{2}$ for comparison\cite{HTC-J}. We utilize our
calculated exchange constants as a function of the wavevector for this
purpose, and not their nearest neighbor fits shown in Table \ref{doping}.
This procedure fully accounts for the long--range effects of the
interactions. Our model demonstrates some differences between the spin--wave
dispersions of NdNiO$_{2}$ and CaCuO$_{2}$. Notably, the peak around $(\frac{%
1}{2},0,0)$ is reduced in NdNiO$_{2}$ compared with CaCuO$_{2}$, as a
consequence of the smaller in--plane exchange couplings, and is largely
unaffected by doping. In contrast, the out--of--plane exchange interactions
strongly depend on doping, which can be seen in the changing dispersion
along $\Gamma Z$. In undoped NdNiO$_{2}$, an out--of--plane $J_{z2}$
dominates over the vanishing nearest neighbor $J_{z1}$. Doping amplifies $%
J_{z1}$ while suppressing $J_{z2}$, resulting in the disappearance of the
valley at $(0,0,1/2)$ in the dispersion. Thus, in contrast with HTSCs, our
calculation of $J$'s here predicts a strongly doping dependent resonance
that could in principle be observed in neutron experiments.

\section{III. TWO--BAND\ MODEL}

\begin{figure}[tbp]
\includegraphics[height=0.45\textwidth,width=0.48\textwidth]{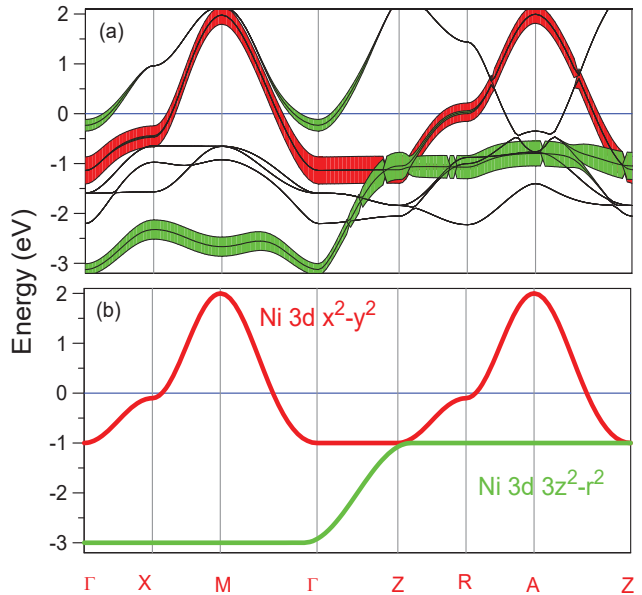}
\caption{(a) Non--magnetic LDA\ band structure of NdNiO$_{2}$ with the
orbital character of Ni-3\emph{d}$_{x^{2}-y^{2}}$ and Ni-3\emph{d}$%
_{3z^{2}-r^{2}}$ states shown in red and green, respectively. (b) The
corresponding two-band tight--binding model\protect\cite{SM}.}
\label{FigTB}
\end{figure}

A minimal model for the electronic structure of NdNiO$_{2}$ that emerges
from the present study should involve Ni--3\emph{d}$_{x^{2}-y^{2}}$ and Ni--3%
\emph{d}$_{3z^{2}-r^{2}}$ orbitals only. Their role has already been
emphasized in many recent works\cite{Norman}--\cite{Spin excitations} and,
as we argue here, their importance is based on sensitivity of magnetic
excitations to the position of various orbitals. The parameters of the model
can be obtained by tracing the orbital character of these states from the
non--magnetic LDA\ calculation. We show this in red for Ni--3\emph{d}$%
_{x^{2}-y^{2}}$ and in green for Ni--3\emph{d}$_{3z^{2}-r^{2}}$ in Fig. \ref%
{FigTB}(a). The derived two--band tight--binding model is illustrated in
Fig. \ref{FigTB}(b). In the large $U$ limit, such model at a quarter filling
by holes (3 electron filling) is expected to exhibit a Mott insulator for 3%
\emph{d}$_{x^{2}-y^{2}}$ band, with the lower Hubbard band placed below 3%
\emph{d}$_{3z^{2}-r^{2}}$ state. Its antiferromagnetic solution in the
Hartree--Fock approximation will result in the band structure very similar
to the LDA+U result shown in Fig. \ref{FigBands}(b), which also assumes that
the $t_{2g}$ states of Ni, although appear in the same energy range, are
apparently irrelevant. According to our LDA+U calculation with $U\gtrsim $4
eV, doping sends the holes primarily to the 3\emph{d}$_{3z^{2}-r^{2}}$ state
promoting interorbital triplets. This is seen in Fig. \ref{FigBands}(b)
where the Fermi level shifting downwards unoccupies the 3\emph{d}$%
_{3z^{2}-r^{2}}$ band in $k_{z}=\pi /c$\ plane which explains doping
dependence of the orbital occupancies shown in Table \ref{doping}.

The described picture should however be contrasted to the genuine strong
correlation effect that prompts to consider an additional hole to be
injected into either Ni $x^{2}-y^{2}$ or $3z^{2}-r^{2}$ orbital resulting
either in an intraorbital singlet or interorbital triplet. This is different
from cuprates, where holes end up in low--lying O 2p band forming
Zhang--Rice singlet states\cite{Zhang-Rice}. Here, it is not the relation of
Hubbard $U$ to the crystal field splitting $\Delta $ between $x^{2}-y^{2}$
and $3z^{2}-r^{2}$ levels but the competition of the Hund's rule $J_{H}$ and 
$\Delta $ which should be examined to understand the origin of the two--hole
state in the doped case\cite%
{Sawatzky,Lechermann,Millis,Multiorbital,Multiorbital2}. To illustrate the
proximity of both (S=0 and S=1) solutions, a simple diagonalization of the 3d%
$^{8}$ shell with U=6 eV and our deduced from Fig. \ref{FigBands}(b) crystal
field splitting $\Delta $=2.2 eV reveals that the lowest energy state is S=0
for $J_{H}<$ 0.9 eV, and S=1 otherwise. This value is well within the range
of generally assumed Hund's rule exchange energies for transition metal
oxides and highlights a delicate balance in extracting the two--hole ground
state configuration.

\begin{figure}[tbp]
\includegraphics[height=0.363\textwidth,width=0.49\textwidth]{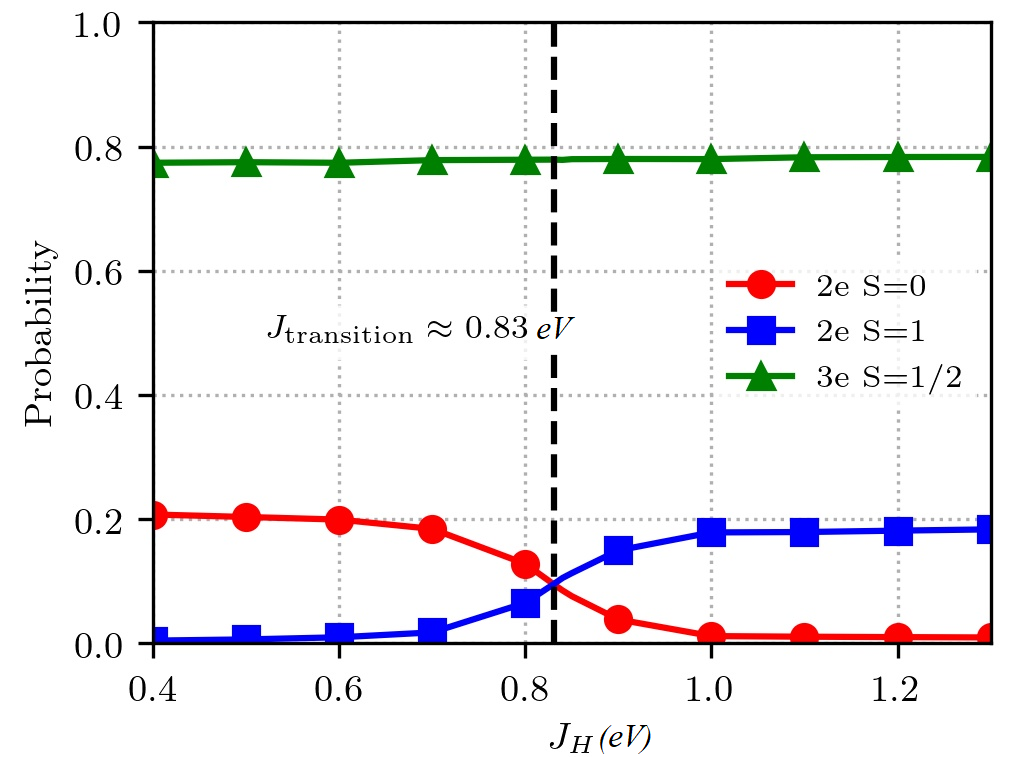}
\caption{Calculated probabilites for the three electron S=1/2 and
two--electron S=0 and S=1 states as a function of Hund's coupling $J_{H}$
using Dynamical Mean Field Theory and Continious Time Quantum Monte Carlo
Method for the two--band model of NdNiO$_{2}$ corresponding to doping by 0.2
holes (filling by 2.8 electrons in the model). An inverse temperature of $%
\protect\beta =40$ is used in the calculations.}
\label{FigQMC}
\end{figure}

Although a number of full--fledge multiorbital LDA+DMFT calculations have
been recently carried out to understand the many--body physics of doped NdNiO%
$_{2}$\cite{Lechermann,Millis,Sergey}, a strong sensitivity of the solution
to the input parameters, such as $J_{H}$, is expected. This has been already
highlighted in the earlier work simulating two semicircular densities of
states with the crystal field splitting as a parameter \cite{Multiorbital2}.
To gain a further qualitative insight, here we study our derived two--band
model using Dynamical Mean Field Theory\cite{DMFT} and Continuous\ Time
Quantum Monte Carlo method\cite{CTQMC}. The parameters $U$=6 eV and $\Delta $%
=2.2 eV are fixed while $J_{H}\ $is adjusted. The undoped case of Ni 3d$^{9}$
S=1/2 state corresponds to the electronic filling equal to 3 in this model,
where we easily recover a paramagnetic Mott insulating state with the gap of
the order of $U$ that opens up in the $x^{2}-y^{2}$ band and with the $%
3z^{2}-r^{2}$ states that remain completely occupied. Doping this model with
0.2 holes (filling by 2.8 electrons) results in finite probability to find
either S=0 or S=1 states in addition to S=1/2 that depends on $J_{H}.$ These
probabilities extracted from the Quantum Monte Carlo simulation are shown in
Fig. \ref{FigQMC} very close to our earlier estimate of 0.9 eV.

\begin{figure}[tbp]
\includegraphics[height=0.695\textwidth,width=0.49\textwidth]{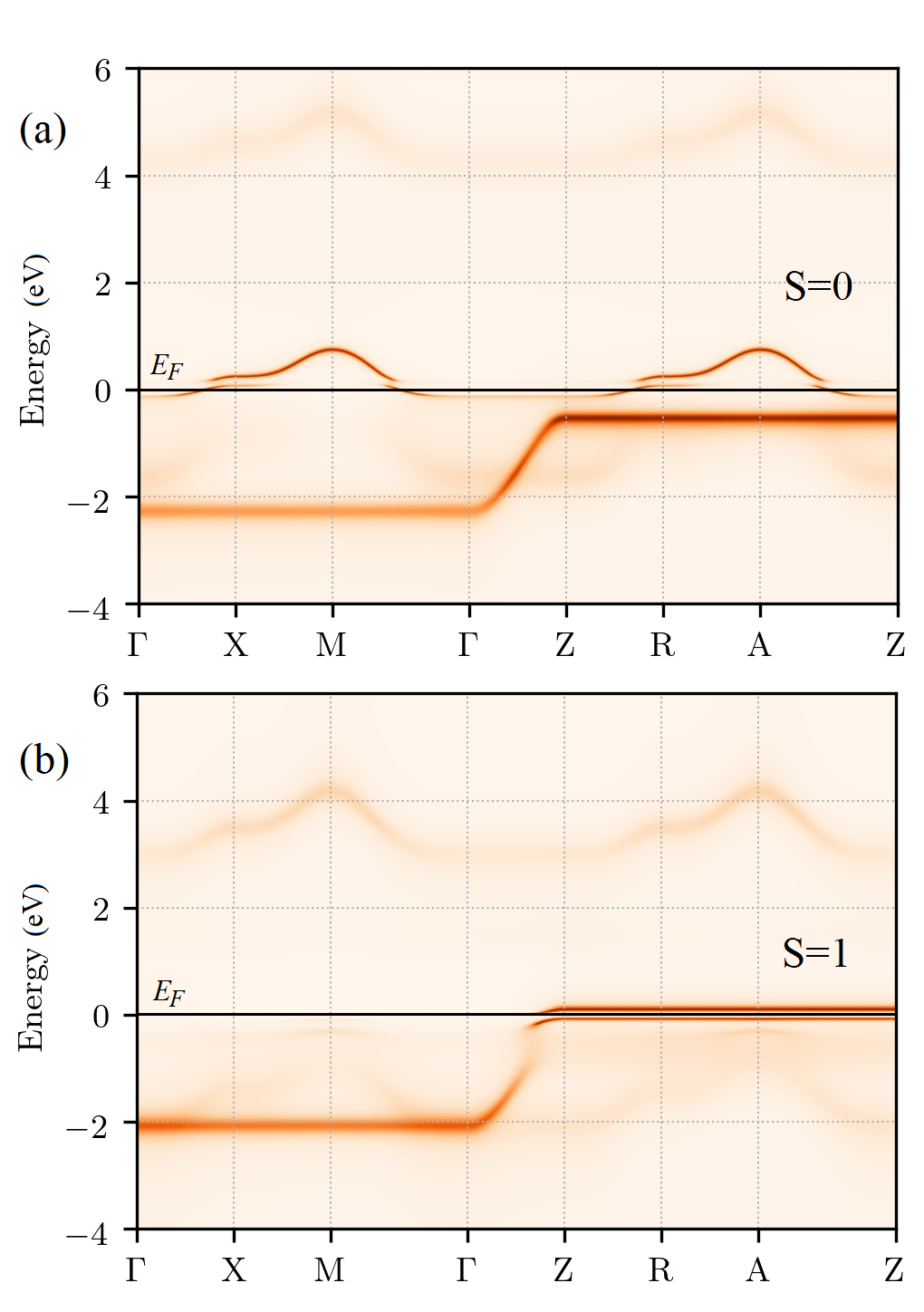}
\caption{Quasiparticle band structures of the two-- band model of NdNiO$_{2}$
obtained by Dynamical Mean Field Theory and Continious Time Quantum Monte
Carlo Method for the doping level of 0.2 holes: (a) Calculation for Hund's
coupling $J_{H}=0.6$ eV that corresponds to S=0 two--hole state; (b)
Calculation for $J_{H}=1$ eV corresponding to S=1 two-hole state. An inverse
temperature $\protect\beta =40$ is used for the calculations.}
\label{FigDMFT}
\end{figure}

Our results for the $\mathbf{k}$--resolved spectral functions are summarized
in Fig. \ref{FigDMFT}, where a comparative study is presented for the two
quasiparticle band structures corresponding to S=0 state ($J_{H}$ is set to
0.6 eV, Fig. \ref{FigDMFT}(a)) and to S=1 state ($J_{H}$ is set to $1$ eV,
Fig. \ref{FigDMFT}(b)). One can see from the calculated spectrum for $J_{H}=$%
0.6 eV that the $3z^{2}-r^{2}$ state remains completely occupied while the
doping primarily affects the $x^{2}-y^{2}$ band which now shows a typical
for DMFT three--peak structure with the two Hubbard bands appearing below
and above the Fermi level and a renormalized quasiparticle band that crosses 
$E_{F}$. The k dispersion for all three features is similar to the original
dispersion of the $x^{2}-y^{2}$ band.

A different picture emerges from the calculation with $J_{H}=1$ eV shown in
Fig. \ref{FigDMFT}(b). In this case, renormalized quasiparticles of the $%
3z^{2}-r^{2}$ character appear at the Fermi level which illustrate the
formation of the interorbital triplet states. A very strong peak in the
quasiparticle density of state is expected to be present at $E_{F}$ due to
the non--dispersive portion of the $3z^{2}-r^{2}$ band within the $ZRA$
plane. At the same time, the $x^{2}-y^{2}$ band does not develop a
three--peak structure and is characterized by the two Hubbard bands as in
the undoped case. A very similar behavior has been already predicted in a
recent work\cite{Multiorbital} where it was termed as the \textquotedblleft
Kondo resonance\textquotedblright\ property, carried by the Ni-$3z^{2}-r^{2}$
character.

Our previous LDA+DMFT calculations \cite{Sergey} performed for $J_{H}=0.95$
eV are in somewhat agreement with this result although the appearance of the
flat band was detected by us earlier only at a higher doping (\symbol{126}%
0.4). The origin of this discrepancy may lie in a more complex interplay
between crystal fields and double counting effects in a self--consistent
multiorbital simulation or in an analytical continuation of the QMC derived
spectral functions resulting in a smaller and/or more broadened spectral
weight as compared to the result of the model. We have additionally checked
the probabilities of various spin states within LDA+DMFT and they are mostly
in line with what we observe in Fig. \ref{FigQMC}.

Since the Hund's coupling $J_{H}$ of 0.8 to 0.9 eV is well within the range
of commonly accepted values, we cannot make a definite conclusion about
whether S=0 or S=1 scenario is realized for doped nickelates. However,
possible future angle--resolved photoemission (ARPES) experiments may
provide important insight since as illustrated by our calculations the
quasiparticle band structure is very different between the two cases.
Furthermore, while ARPES\ spectra in the hole--doped HTSCs show
waterfall--like behavior\cite{Waterfall}, we do not expect waterfalls to
appear here due to a lack of oxygen states at energies close to $E_{F}$ and
associated physics responsible for the formation of the low energy states%
\cite{Gordi}.

\section{IV. CONCLUSION}

In conclusion, using the LDA+U method, we have calculated magnetic exchange
interactions for the doped NdNiO$_{2}$ novel superconductor. We find that
the parent compound is mostly two--dimensional, with large nearest neighbor
in--plane, and small out--of--plane exchange interactions. Upon doping, the
out--of--plane coupling $J_{z1}$ was found to increase dramatically, while
the in--plane $J_{1}$ is almost unchanged. To clarify the origin of these
trends, we analyzed the symmetry of the holes induced by doping which were
found to be primarily of the 3\emph{d}$_{3z^{2}-r^{2}}$ character promoting
the formation of interorbital triplet as Ni 3d$^{8}$ ground state
configuration. We also investigated the role of the Nd--5\emph{d} states,
which contribute substantially to the Fermi surface of NdNiO$_{2}$. Shifting
this band upward using a constrained--orbital--hybridization method has
little effect on exchange interactions, which leads us to conclude that Nd--5%
\emph{d} states have negligible effect on the spin fluctuations and the
superconductivity in NdNiO$_{2}$. A minimal two--band model with active Ni--3%
\emph{d}$_{x^{2}-y^{2}}$ and Ni--3\emph{d}$_{3z^{2}-r^{2}}$ orbitals has
been further studied with DMFT to reveal an underlying Mott insulating state
which upon doping selects either S=0 and S=1 two--hole states depending on
the Hund's coupling in the range of its commonly accepted values 0.8 to 0.9
eV. Should S=1 state be valid, we\ rely on our LDA+U\ result to predict that
upon doping the spin susceptibility gains three dimensionality as it gets
enhanced along $\Gamma Z.$ This can be readily observed in neutron
experiments and can be one way to probe the two--hole configuration. We also
rely on our DMFT result to predict a formation of a strong quasiparticle
peak at the Fermi level detectable by ARPES experiments. A small anisotropy
in $H_{c2}$ was indeed discovered very recently \cite{H.H.Wen} illustrating
the three--dimensional nature of NdNiO$_{2}$ which starkly contrasts with
the two--dimensional superconductivity in HTSCs. At the same time, most
recent x--ray absorption spectroscopy (XAS) and resonant inelastic x-ray
scattering (RIXS) experiments are found to be consistent with a d$^{8}$ spin
singlet state \cite{Devereaux}. These results should be important in future
studies of nickelate superconductors.

\section{\textbf{Acknowledgements}}

X.W. is supported by the NSFC (Grants No. 11834006, No. 11525417, No.
51721001, and No. 11790311), National Key R\&D Program of China (Grants No.
2018YFA0305704 and No. 2017YFA0303203) and by 111 Project. X.W. also
acknowledges the support from the Tencent Foundation through the XPLORER
PRIZE . V.I, G.R. and S.Y.S are supported by NSF DMR Grant No. 1832728. I.L.
acknowledges support by the Russian Foundation for Basic Research (Project
No. 18-32-20076). The DMFT electronic structure calculations were supported
by the state assignment of Minobrnauki of Russia (theme \textquotedblleft
Electron\textquotedblright\ No. AAAA-A18-118020190098-5).

\end{document}